\documentclass[useAMS,usenatbib,usegraphicx]{mn2e}
\usepackage{times}
\usepackage{lscape}
\usepackage{amssymb}
\usepackage{color}

\def\cm3{$\rm cm^{-3}$}

\def\n0{$\rm n_{0}$}
\def\B0{$\rm B_{0}$}

\def\mc{$\mu$m}

\def\L12{L$_{12\mu m}$~}
\def\F12{F$_{12\mu m}$~}

\def\fe2{[Fe\,{\sc ii}]}
\def\s3{[S{\sc iii}]}
\def\h2{H$_{2}$}

\def\F{$F_{\lambda}$}

\def\paasp{{\sc p{\tt a}asp}}

\title[\paasp]{Panchromatic Averaged Stellar Populations: \paasp}

\author[Riffel et al.]{R. Riffel$^{1}$\thanks{E-mail:
riffel@ufrgs.br}, C. Bonatto$^1$, R. Cid Fernandes $^2$, M. G. Pastoriza$^{1}$ and E. Balbinot$^{1}$
\\$^{1}$Departamento de Astronomia, Universidade Federal do Rio Grande do Sul. Av. Bento Gon\c calves 9500, Porto Alegre, RS, Brazil.
\\$^2$ Departamento de F\'{\i}sica - CFM - Universidade Federal de Santa Catarina, Florian\'opolis, SC, Brazil}

\begin{document}

\date{}

\maketitle


\begin{abstract}
{\bf We study how the spectral fitting of galaxies, in terms of light fractions
derived in one spectral region translates into another region, by using results 
from evolutionary synthesis models. In particular, we examine propagation dependencies 
on Evolutionary Population Synthesis (EPS, {\sc grasil}, {\sc galev}, Maraston and 
{\sc galaxev}) models, age, metallicity, and stellar evolution tracks over the 
near-UV---near infrared (NUV---NIR, 3500\AA\ to 2.5\mc) spectral region.} Our main results are:
as expected, young ($t \lesssim$ 400 Myr) stellar population fractions derived in the optical cannot be directly 
compared to those derived in the NIR, and vice versa.  In contrast, intermediate to 
old age ($t \gtrsim$ 500 Myr) fractions are similar over the whole spectral region studied.
The metallicity has a negligible effect on the propagation of the stellar population fractions derived from 
NUV --- NIR. The same applies to the different EPS models, but restricted to the range between 3800 \AA\ and 9000 \AA. However, a discrepancy 
between {\sc galev}/Maraston and {\sc grasil}/{\sc galaxev} models occurs in the NIR.
Also, the initial mass function (IMF) is not important for the synthesis propagation.
Compared to {\sc starlight} synthesis results, our propagation predictions agree at 
$\sim$95\% confidence level in the optical, and $\sim$85\% in the NIR.  
{\bf In summary, spectral fitting} performed in a restricted spectral range should not be directly propagated from 
the NIR to the UV/Optical, or vice versa. We provide equations and an on-line form 
({\bf Pa}nchromatic {\bf A}veraged {\bf S}tellar {\bf P}opulation - \paasp) to be used for this purpose.
\end{abstract}

\section{Introduction}

A key issue in modern astrophysics is to understand how galaxies form and evolve, and the study of
the stellar population and star formation history (over a wide wavelength range) may provide clues to
the dominant mechanism. For example, two main 
scenarios are proposed to explain the formation of galaxies, one is the tidal torque theory, which suggests an
initial collapse of gas at very high redshift \citep[e.g.][]{eggen62,withe84}, and another is 
associated with galaxy mergers at intermediate to high redshifts 
\citep[see][for example]{searle78,hammer05,hammer07,hammer09}. Both produce distinct 
signatures in the resulting stellar populations. 

As we still cannot access the light of individual stars in almost any 
galaxy beyond the local group, the way widely used is to 
study the stellar populations by the integrated light of the whole galaxy or a significant fraction of it,
by means  of long-slit spectroscopy.
The usual approaches for studying unresolved stellar populations are either by means of empirical population 
synthesis \citep{bica87,bica88,bica91,charles98,charles2000,cid98,schmitt96} or  
Evolutionary Population Synthesis (EPS) models \citep{bc03,maraston98,maraston05,vazdekis99,silva98,kotulla09}, 
as well as a combination of both techniques \citep{cid04,cid05,riffel07,riffel08,riffel09}. By using them, we 
can infer the age and metallicity distributions of the stellar populations that make up 
a galaxy's spectral energy distribution (SED).

The study of unresolved stellar populations in galaxies, ranging from star-forming 
to ellipticals, as well as in composite objects and active galaxies, is a 
common approach in the Near-UV (NUV) and  optical bands \citep{bica88,schmitt96,cid98,
rosa98,charles98,charles2000,raimann03,rosa04,cid05,angela08,mauro08,mauro09}. On the other 
hand, addressing unresolved stellar populations 
in the near infrared (NIR) is less common, starting $\sim$3 decades 
ago \citep[e.g.]{rieke80},  but which seems to be in increasing expansion \citep{orig93,oliva95,engelbracht98,origlia00,lancon01,davies06,davies07,davies09,riffel07,riffel08,riffel09,riffel10}.

Very recently, \citet{chen10} have shown that using different EPS models in the optical leads to different 
stellar population results. So, it is not reasonable to directly compare stellar populations estimated 
from different EPS models. They suggest that to get reliable results, one should use the same 
EPS models to compare different samples. As related issues, 
can one compare the light-fraction results derived, with the same base of elements, in one spectral 
region to another? Are the population fractions derived in the optical the same in the NUV/NIR? What is the effect of the 
normalisation point used in the synthesis? Does the choice of elements that compose the 
base (i.e. age, metallicity and initial mass function - IMF) produce different results along 
the whole spectrum?

The answer to the above questions would be easy {\bf if stellar population synthesis were 
based on} a more physical parameter, like mass-fractions. However, given the non-constant 
stellar mass-to-light ($M/L$) ratio, mass-fractions have a much less direct relation with 
the observables than light-fractions. In this context, we feel motivated to carry out a 
detailed study of the panchromatic averaged stellar population (\paasp) components over 
the 3500\AA\ to 2.5\mc\ spectral region, commonly used for stellar population synthesis.  
{\bf It should be clear that \paasp\ is not a tool for performing stellar population 
synthesis. Instead, it is specifically designed for translating the spectral fitting of 
galaxies (in terms of light fractions) derived in one spectral domain into another. }
This paper is structured as follows: The  EPS models used are described in Sect.~\ref{epsmodels}. 
The methodology and results are presented in Sect.~\ref{method}.
Results are discussed in Sect.~\ref{discussion}. The final remarks are given in Sect.~\ref{finalremarks}.

\section{The adopted EPS models }\label{epsmodels}

In this section we describe the EPS models used in what follows. The four models: 
{\sc grasil}, {\sc galev}, Maraston and {\sc galaxev}, have been selected because they have 
a spectral coverage from the NUV to the NIR region ($\sim$3500\AA\ - 2.5\mc) and are widely 
used. A brief description of them is made below. 
Details can be found in the original papers cited below, as well as in \citet{chen10}.

\subsection{{\sc grasil}}

The GRAphite and SILicate - {\sc grasil}\footnote{Available at: http://adlibitum.oats.inaf.it/silva/grasil/grasil.html} - code, developed 
by \citet{silva98}, is a chemical evolution code that 
follows the star formation rate, metallicity and gas fraction, which are basic ingredients for 
stellar population synthesis. The latter is performed with a grid of integrated spectra 
of simple stellar populations (SSPs) of different ages and metallicities, in which the effects of dusty 
envelopes around asymptotic giant branch (AGB) stars are included,  but not the AGB energetics \citep{bressan98,bressan02}.
The models consider four initial mass functions (IMFs), \citet{kurucz92,kennicutt94,scalo86,miller79}. 
\citet{kurucz92} atmosphere models, for population synthesis, and Padova tracks \citep{bertelli94} 
are also considered. Moreover, SSPs of {\sc grasil} cover a spectral range 
from 91\AA\ up to 1200\mc. Further details can be found in Laura Silva PhD 
thesis\footnote{see: http://adlibitum.oats.inaf.it/silva/laura/laura.html}.

\subsection{{\sc galev}}

The GALaxy EVolution - {\sc galev}\footnote{http://www.galev.org} -  evolutionary synthesis models describe the evolution of stellar populations
in general, from star clusters to galaxies, both in terms of resolved stellar
populations and integrated light properties \citep[][]{kotulla09}.
According to the authors, the code considers both the chemical evolution of the gas and the spectral evolution 
of the stellar component, allowing for 
a chemically consistent approach. Thus, some SSPs provided by {\sc galev} show a emission line spectra. 

The SSP models provided by {\sc galev} cover 5 metallicities and 4000 ages ($\rm 0.02 \leq Z/Z_{\odot} \leq 2.5$ and $\rm 4\, Myr \leq t \leq 16 Gyr$). They 
are based on the spectra from BaSeL spectral library \citep{lejeune97,lejeune98,westera02}, originally
based on the \citet{kurucz92} library. They also have 3 different IMFs \citep{salpeter55,kroupa01,chabrier03}
and use the theoretical isochrones from the Padova team \citep[e.g.][]{bertelli94,schulz02}. 
The wavelength coverage spans the range from 90\AA\ to 160\mc, with a spectral resolution of 20\AA\ in
the NUV-optical, and 50-100\AA\ in the NIR. They also include the thermally- pulsating asymptotic giant
branch (TP-AGB) phase provided by the Padova tracks \citep{schulz02,girardi02}. It is also worth mentioning at this point that the TP-AGB stars account 
for 25 to 40\% of the bolometric light of an SSP, and for 40 to 60\% of the light emitted in the K-band \citep[e.g.][and references 
therein]{schulz02,maraston05}.  In addition, contrary to  \citet[][see below]{maraston05}, {\sc galev} models do not include empirical 
TP-AGB spectra. For a detailed description of {\sc galev} see \citet{kotulla09}.

\subsection{Maraston models (M05)}

The Maraston EPS models\footnote{Available at: http://www-astro.physics.ox.ac.uk/$\sim$maraston/} (hereafter M05) 
are being developed by Claudia Maraston since 1998 \citep{maraston98} with an update in 2005 \citep{maraston05}. They 
are based on the fuel consumption theorem and include a proper treatment of the TP-AGB phase. According to these models, the effects of 
TP-AGB stars in the NIR spectra are unavoidable. The M05 models, by including empirical spectra of oxygen-  and carbon-rich stars \citep{lw00}, 
can predict the presence of NIR absorption features such as the 1.1\mc\ CN band \citep{riffel07,riffel08,riffel09}, whose detection can be taken as 
an unambiguous evidence of a young to intermediate age SP.  The models have been used by our team to study stellar populations in active galactic 
nuclei and starburst galaxies \citep[][for CN, see also Ramos-Almeida et al. 2008 and Dottori et al. 2005]{riffel07,riffel08,riffel09,riffel10} 
as well as in the age dating of massive galaxies at high redshift \citep{maraston06,van06,rodig07,cimatti08}.

M05 models span a range of 6 different metallicities (${\rm \frac{1}{200}} \leq \frac{Z}{Z_{\odot}} \leq \rm 3.5$)
with ages distributed from 1 Myr to 15 Gyr according to a grid of 67 models (Note that the full age grid is not 
available for all metallicities,  M05).  The IMFs considered are: \citet{salpeter55} and \citet{kroupa01}. The stellar spectra were also taken
from the BaSel library. The spectral range is from  91\AA\ to 160\mc, with a spectral 
a resolution of 5-10\AA\ up to the optical region, and 20-100\AA\ in the NIR.
 In \citet{maraston98,maraston05} models, the TP-AGB contributes with 40\% to the bolometric flux, but rising to 80\% when only 
the $K$-band is considered.

\subsection{{\sc galaxev}}

{\sc galaxev}\footnote{Available at: http://www.cida.ve/$\sim$bruzual/bc2003} is a widely used library of evolutionary stellar population 
synthesis models. It is  computed with the isochrones synthesis code of \citet[][also known as BC03]{bc03} 
The spectral coverage of this library is from 91 \AA\ up to 160\mc, with a resolution of 3 \AA\ 
between 3200 and 9500 \AA, and a lower resolution elsewhere. Ages range from  $\rm 1 \times 10^5$ up to $\rm 2 \times 10^{10}$ yr, 
for a wide range of metallicities (${\rm \frac{1}{200}} \lesssim \frac{Z}{Z_{\odot}} \lesssim \rm 2.5$).
These models use the STELIB/BaSeL libraries \citep[see][and references therein]{lejeune97,lejeune98,westera02} as well as the STELIB/Pickles libraries
\citep{pickles98}. {\sc galaxev} allows the use of two IMFs \citep{chabrier03,salpeter55} and 3 stellar 
evolution tracks: Geneva \citep{schaller92},
Padova\,94 \citep{alongi93,bressan93,fagotto94a,fagotto94b,girardi96} and Padova\,00 \citep{girardi00}.  These models do not include the TP-AGB phase.

\section{Methodology and Results}\label{method}

First, we investigate the dependence of the stellar population components
on the normalisation point, from the NUV to the NIR. To do this,
we select spectral regions free from emission/absorption lines to be used as normalisation 
points, \F. They are: 3800\AA\  4020\AA\  4570\AA\  5300\AA\  5545\AA\  5650\AA\  5800\AA\  5870\AA\  6170\AA\  
6620\AA\  8100\AA\  8815\AA\  9940\AA\  1.058\mc\ 1.223\mc\ 1.520\mc\ 1.701\mc\ 2.092\mc\ 2.19\mc\ \citep{bica88,mauro08,mauro09,cid04,fatima01,riffel08}.

In addition, we select SSPs with ages: 0.005, 0.025, 0.050, 0.1, 0.2, 0.3,0.4 0.5,0.6, 0.7, 0.8, 0.9, 1.0, 2.0, 5.0 and 13 Gyr as representative 
of the stellar populations observed in galaxies. As a first exercise we combine two components: 13\,Gry (the old population) 
and one of the other SSPs representing the ``young" population. The combination was made by summing up, along the whole spectral 
range ($\sim$3500\AA\ to 2.5\mc), increasing fractions of the ``young" component from 1 to 100\%, according to 
\begin{equation}
F=\left(1-f\right) F_{\rm y} + f F_{\rm o};
\label{eqfrac2}
\end{equation}
where $f$ is the fractional flux, which we vary in steps 
of 0.01, $F_{\rm y}$ is the flux of spectrum of the ``young" component, for each $\lambda$ between $\sim$3500\AA\ to 2.5\mc, 
normalised to unity at 5870\AA\ for the optical, or 
at 1.223\mc\ for the NIR, and $F_{\rm o}$ is the flux of the 13\,Gyr spectrum also normalised at the same points.  Note 
that the normalisation of the spectra is done by dividing their fluxes by that of the normalisation point (5870\AA\ or 1.223\mc),
after the computations are done. In addition, by normalising the spectra we are dealing with light-fractions, which are directly related to the observations.
$\rm F$ is the resulting flux of the combined SSPs (young + old) at a specific $\lambda$. The 
averaged stellar population components spread over all \F\ were derived by:
\begin{equation}
f_{\rm y} = \frac{(1-f) F_{\rm y}}{F}\ {\rm and}\ f_{\rm o} = \frac{f F_{\rm o}}{F}\ 
\label{eqfrac}
\end{equation}

\begin{figure*}
\centering
\includegraphics[scale=0.95]{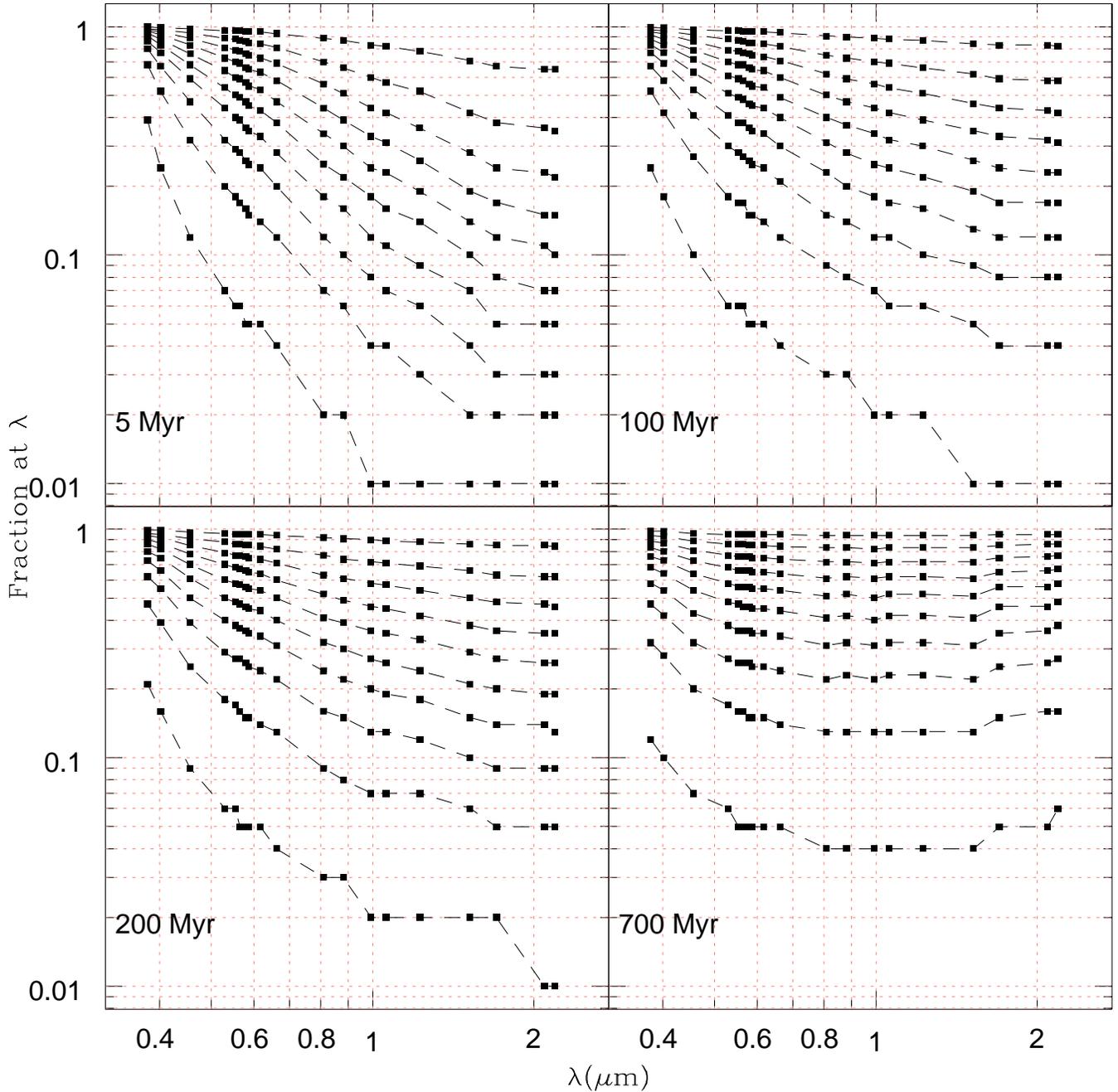}
\caption{Light-fraction at $\lambda$ for the stellar population components. The curves result from Eq.\ref{eqfrac} with increasing 
fractions of 10\% of the ``young" component from bottom to top. In each panel we start with 5\% of the flux of the ``young" SSP 
+ 95\% of the 13\,Gyr SSP, and finish with 95\% of the young + 5\% of the old component. The ages of the ``young" component 
are on the labels. The normalisation point for the combination was 5870\AA.  We use the M05 models as reference. See text for more details.}
\label{fractionsOPT}
\end{figure*}

\begin{figure*}
\centering
\includegraphics[scale=0.95]{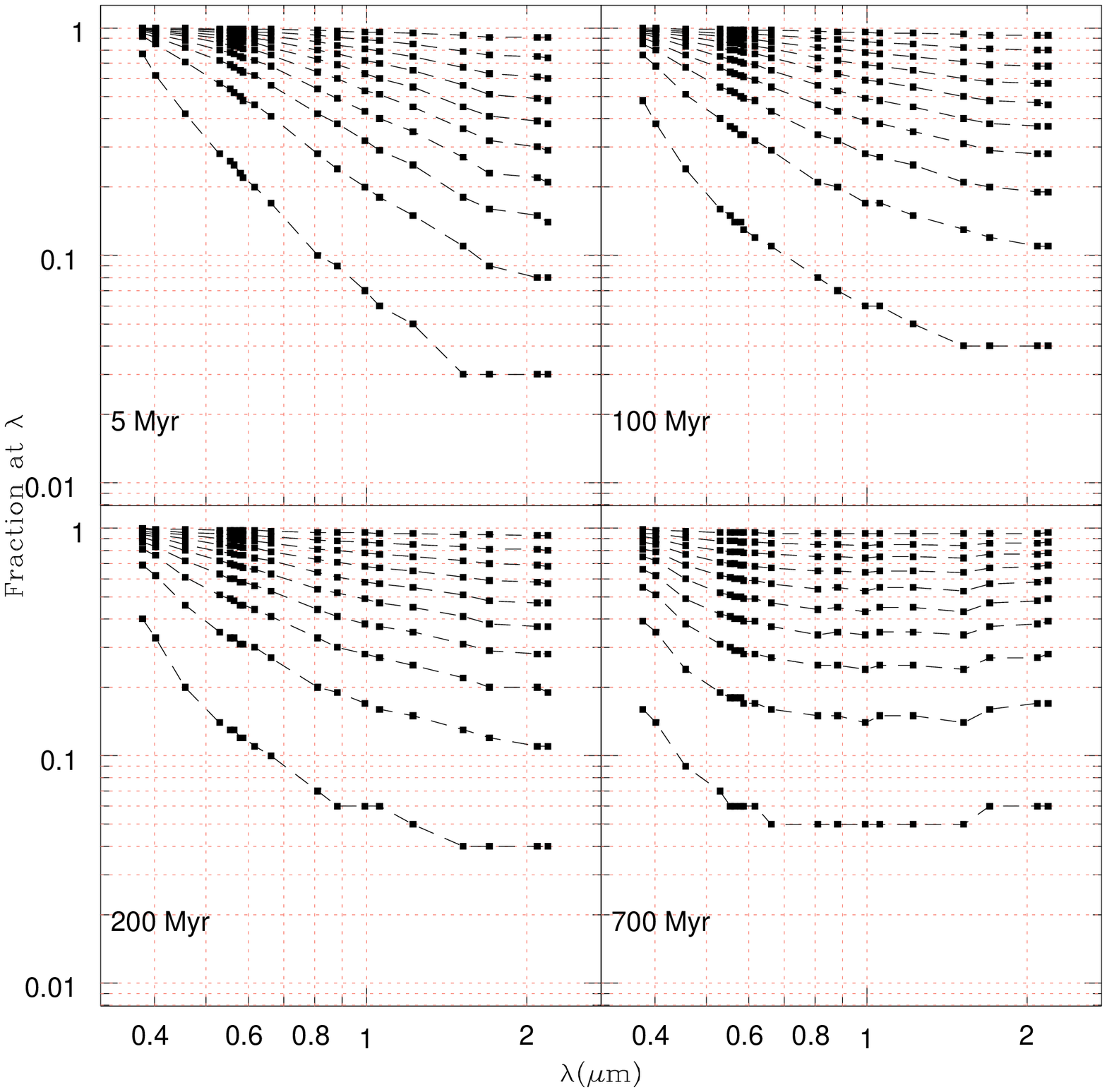}
\caption{Same as Fig.~\ref{fractionsOPT} for the normalisation point at 1.223\mc.}
\label{fractionsNIR}
\end{figure*}

Eq.\ref{eqfrac} represents the young and old light fractions for different \F, this is what we call fraction at $\lambda$.  The result of such 
process is summarised in Figs.~\ref{fractionsOPT} and \ref{fractionsNIR}.  In these figures we show the sum 
of a ``young" component with a 13\,Gyr SSP (the old component) in steps of 10\%. In practice we start with 5\% of the flux of the ``young" SSP 
+ 95\% of the flux of the 13\,Gyr SSP, and finish with 95\% of the young + 5\% of the old component. This procedure is done over all $\lambda$s 
between the NUV and NIR.  It is worth mentioning that Figs.~\ref{fractionsOPT} and \ref{fractionsNIR}, show M05 EPS models as reference 
(Sec.~\ref{discussion}).
These figures suggest that one cannot directly
compare the light fractions of the young component ($t \lesssim$ 400 Myrs) derived in the optical with those obtained in the NIR, and 
vice versa (i.e. 77\% of the 5\,Myr population at 3800\AA\ represents only 5\% at 1.223\mc). However, the intermediate to old components
 ($t \gtrsim$ 500 Myrs) can be directly compared between different wavelengths. This
does not occur when dealing with mass fractions, since the age derived by the mass 
fraction is a more physical parameter, but has a much less direct relation with the observables, depending strongly on the $M/L$ ratio, which is not 
constant. Thus, the light fraction can be taken as an direct observable parameter and use
Eqs.~\ref{eqfrac2} and \ref{eqfrac}, to propagate the results over other spectral regions (see Appendix~\ref{appen}).

However, the stellar population of galaxies is not so simple as two components. Therefore, we have divided our SSPs into 3 population 
vectors $x_y$=0.005\,Gyr;  $x_i$=0.025...1.0\,Gyr and  $x_o$=13\,Gyr. To investigate the effect of adding one more component to the above exercise, we
combined three population vectors according to:

\begin{eqnarray}
F' = \frac{1}{100} \left(\gamma F_{\rm y} + \delta F_{\rm i} + \eta F_{\rm o}\right);\\ 
{\rm with}\ \ \ \gamma + \delta + \eta=100 \nonumber
\end{eqnarray}
where, $\gamma$, $\delta$ and $\eta$ are the fractional fluxes from 0\% to 100\%, which are varied in steps of 1\%; $F_{\rm y}$, $F_{\rm i} $ 
and $F_{\rm o}$ are the normalised fluxes (at $\lambda$=5870\AA\ or $\lambda$ 1.223\mc) of the $x_{\rm y}$, $x_{\rm i}$ and $x_{\rm o}$ population vectors, respectively. 
$\rm F'$ is the resulting flux of the combined SSPs (young + intermediate + old) at a specific $\lambda$.

The averaged stellar population components distributed over all $\lambda$s were derived similarly to Eq.~\ref{eqfrac}: 
\begin{equation}
f'_{\rm y} = \frac{\gamma F_{\rm y}}{F'};\ \ \ f'_{\rm i} = \frac{\delta F_{\rm i}}{F'}\ \ \ {\rm and}\ \ \ f'_{\rm o} = \frac{\eta F_{\rm o}}{F'}\ 
\label{eqfrac3}
\end{equation}

\begin{figure*}
\begin{minipage}[b]{0.45\linewidth}
\includegraphics[scale=0.45]{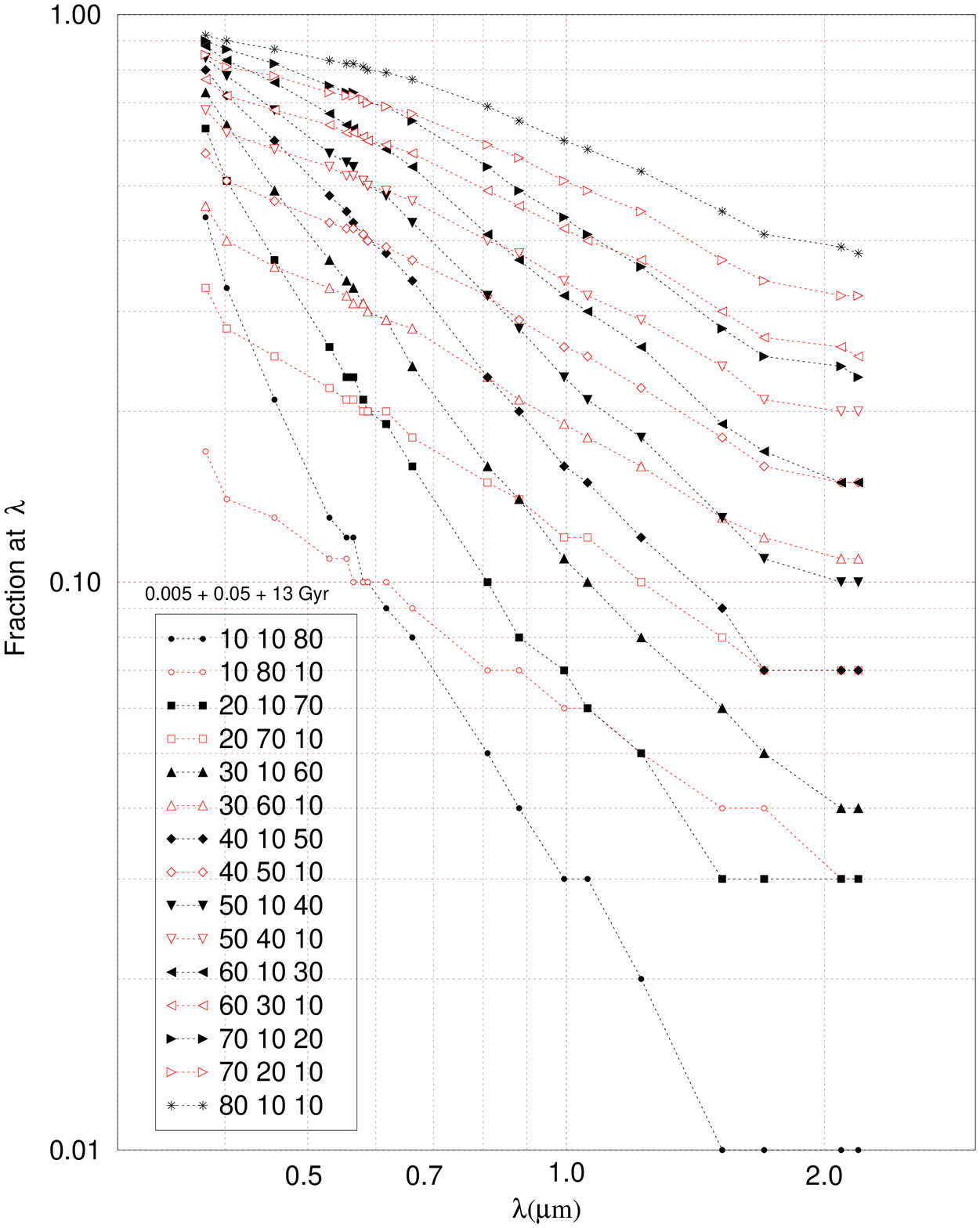}
\end{minipage}
\hfill
\begin{minipage}[b]{0.45\linewidth}
\includegraphics[scale=0.45]{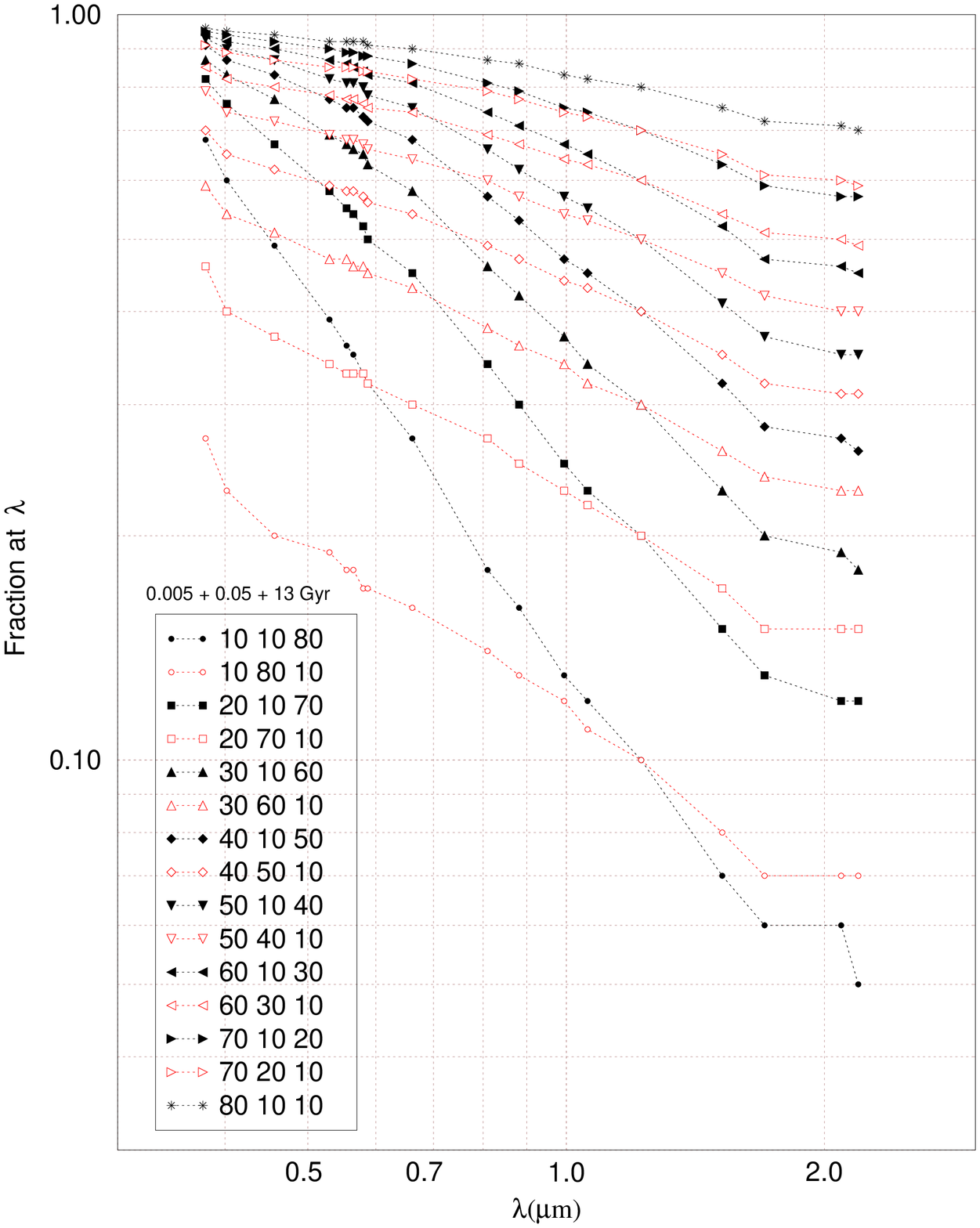}
\end{minipage}
\caption{Light-fraction at $\lambda$ for the stellar population components.The curves result from Eq.\ref{eqfrac3} with increasing 
fractions from bottom to top in each panel. The ages and the percentage fluxes are on the labels. The 5 Myr fraction increases from bottom to top.
The normalisation points are 5870\AA\ (left panel) and  1.223\mc\ (right panel). We use the M05 models as reference.}
\label{3compsNIRa}
\begin{minipage}[b]{0.45\linewidth}
\includegraphics[scale=0.45]{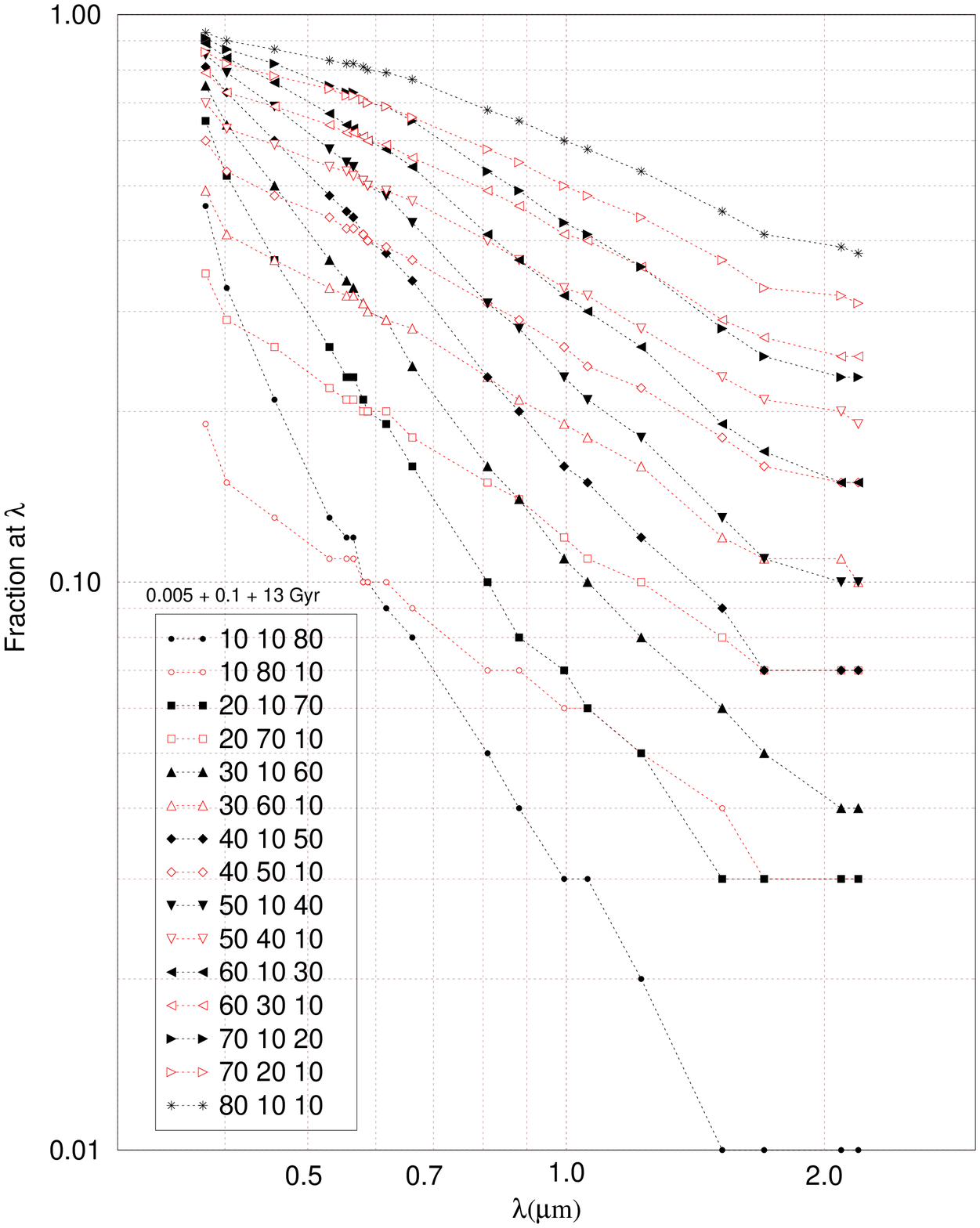}
\end{minipage}
\hfill
\begin{minipage}[b]{0.45\linewidth}
\includegraphics[scale=0.45]{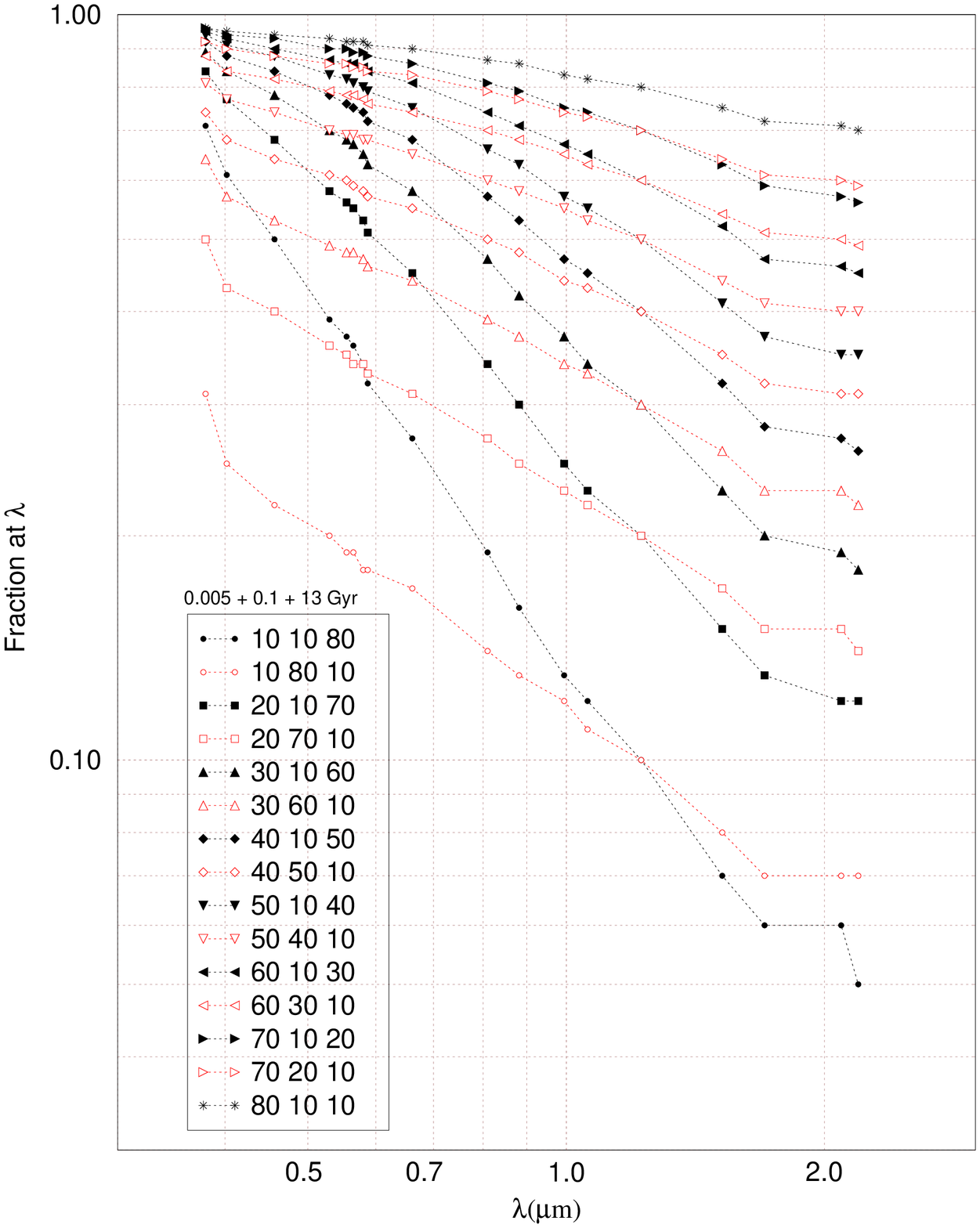}
\end{minipage}
\caption{Same as Fig.~\ref{3compsNIRa} for a different set of ages.}
\label{3compsNIRb}
\hfill
\end{figure*}

\begin{figure*}
\begin{minipage}[b]{0.45\linewidth}
\includegraphics[scale=0.45]{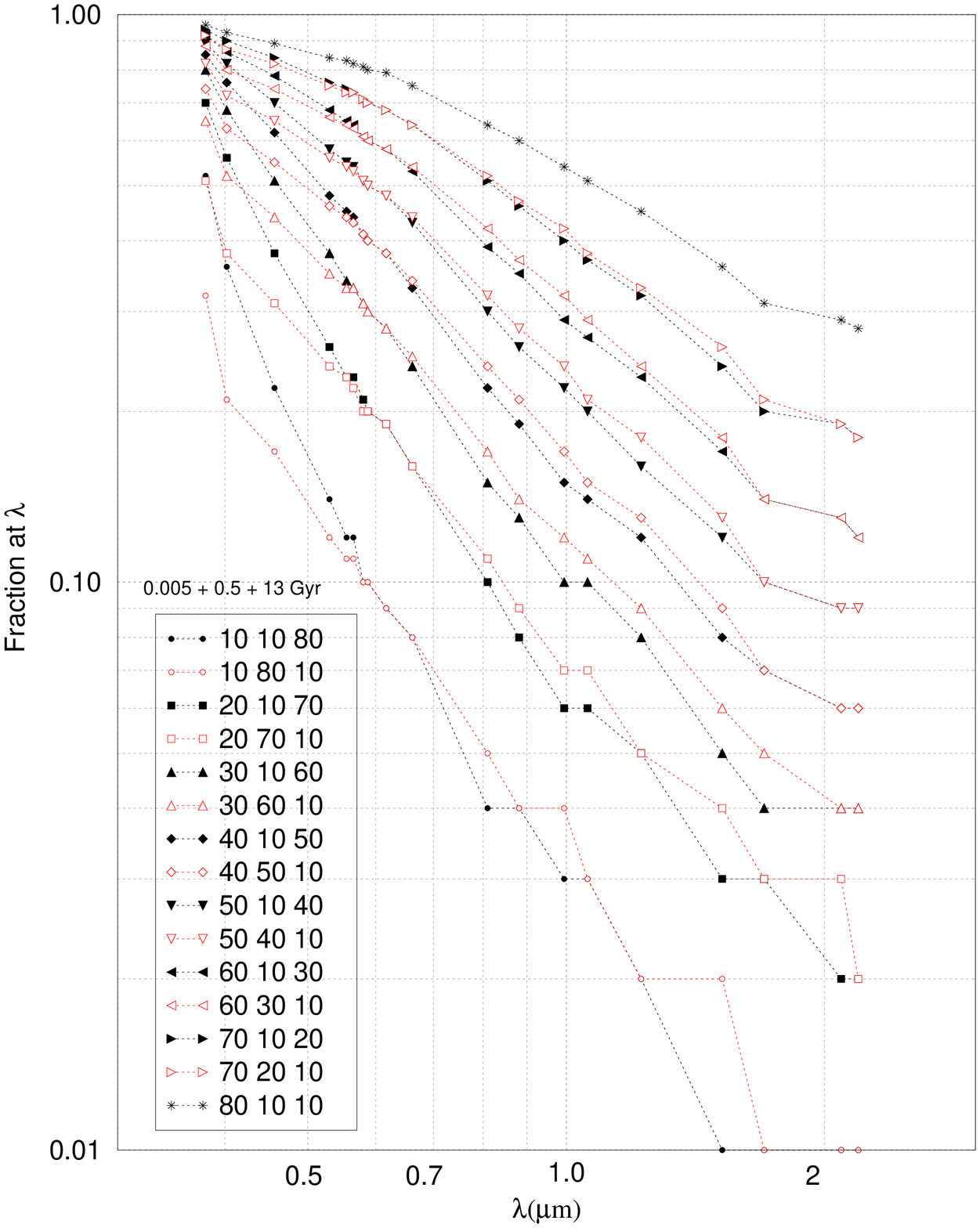}
\end{minipage}
\hfill
\begin{minipage}[b]{0.45\linewidth}
\includegraphics[scale=0.45]{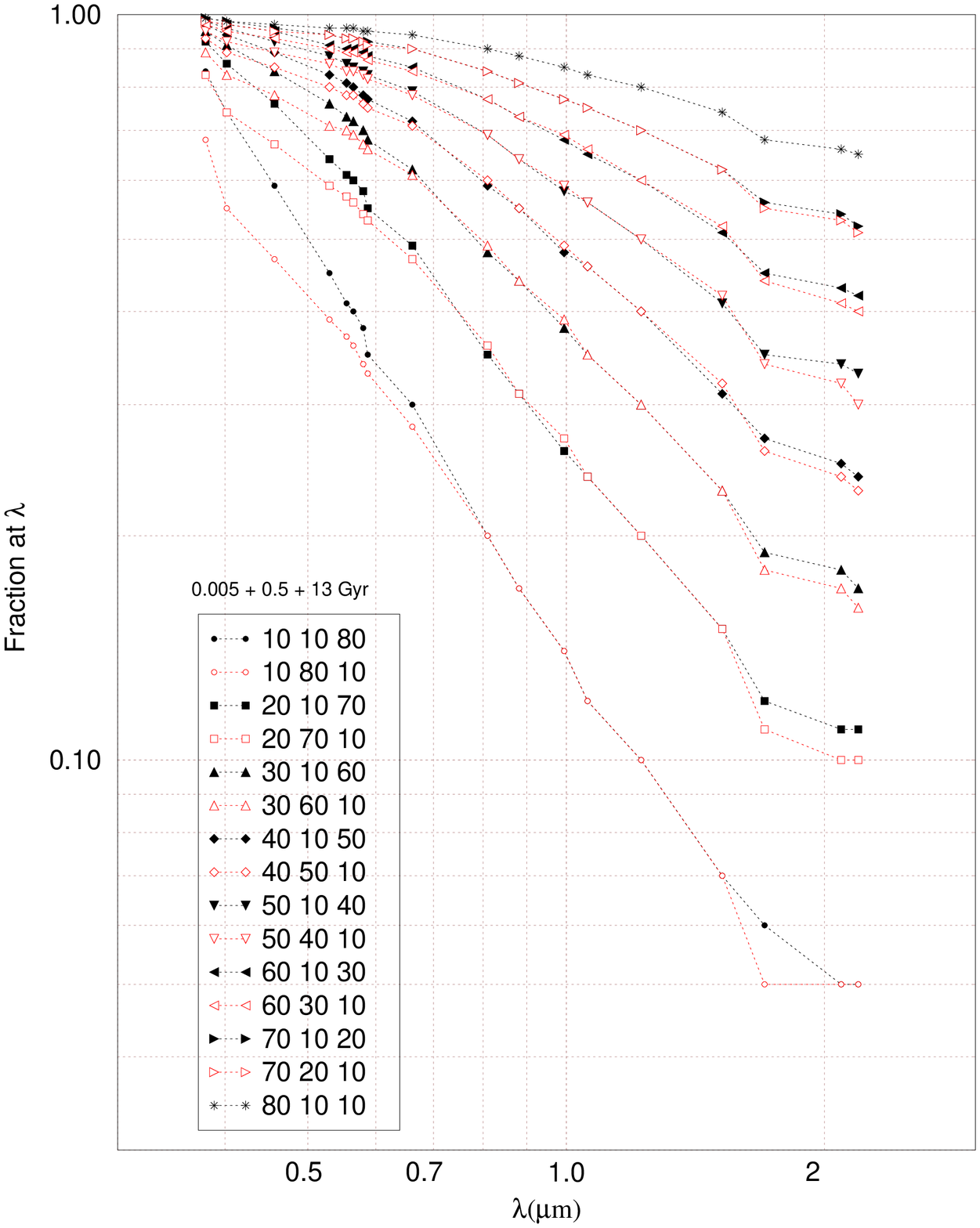}
\end{minipage}
\caption{Same as Fig.~\ref{3compsNIRa} for a different set of ages.}
\label{3compsNIRc}
\hfill
\end{figure*}

We show the result of such combination in Figs.~\ref{3compsNIRa} to \ref{3compsNIRc}. It is clear that even with three 
population vectors, the results 
derived in one wavelength are not the same as in other $\lambda$s. In addition, Fig.~\ref{3compsNIRc} reinforces the fact that in the case of only intermediate to old components, 
the population fractions derived in one wavelength are nearly the same as in other $\lambda$s (i.e. the values
 derived in the optical can be used in the NIR).

\section{Discussion}\label{discussion}

The determination of mean ages and metallicities of galaxies is important
for models of galaxy formation and evolution. The techniques for dating unresolved stellar populations in 
local galaxies have focused on colours or spectroscopic indices measured in  the NUV and optical
\citep[e.g.][and references therein]{bica88,worthey94,bica96,charles96,charles98,trager00a,trager00b,maraston00,thomas05,mauro08}.
However, even small mass fractions of young stars added to an old population can affect the NUV and optical age determinations 
significantly, making the galaxy appear young, which leads to a further degeneracy 
between mass fraction and age \citep{thomas06,serra07}. These problems may be solved for intermediate-age stellar 
populations looking in NIR wavelengths. In addition, the detection of young stellar populations in the NIR requires a 
hard work  \citep{riffel10}. Nevertheless, it is clear from Fig.~\ref{fractionsNIR} that even a small fraction of a 5\,Myr 
population ($\sim$ 5\%) detected in the NIR may be responsible for almost all the light observed in the NUV ($\sim$ 70\%). 

Clearly, synthesis results should not be directly propagated from the NIR to the NUV/Optical, or vice versa. Instead, 
Eqs.~\ref{eqfrac} to ~\ref{eqfrac3} should be used for this purpose. 
To help with such a comparison we have created 
an on-line form, the {\bf Pa}nchromatic {\bf A}veraged {\bf S}tellar {\bf P}opulation: 
\paasp\footnote{available at: http://www.if.ufrgs.br/$\sim$riffel/software.html} and make available for download the tables
with the results of the above equations (see Appendix~\ref{appen}).

\begin{figure*}
\centering
\includegraphics[scale=0.95]{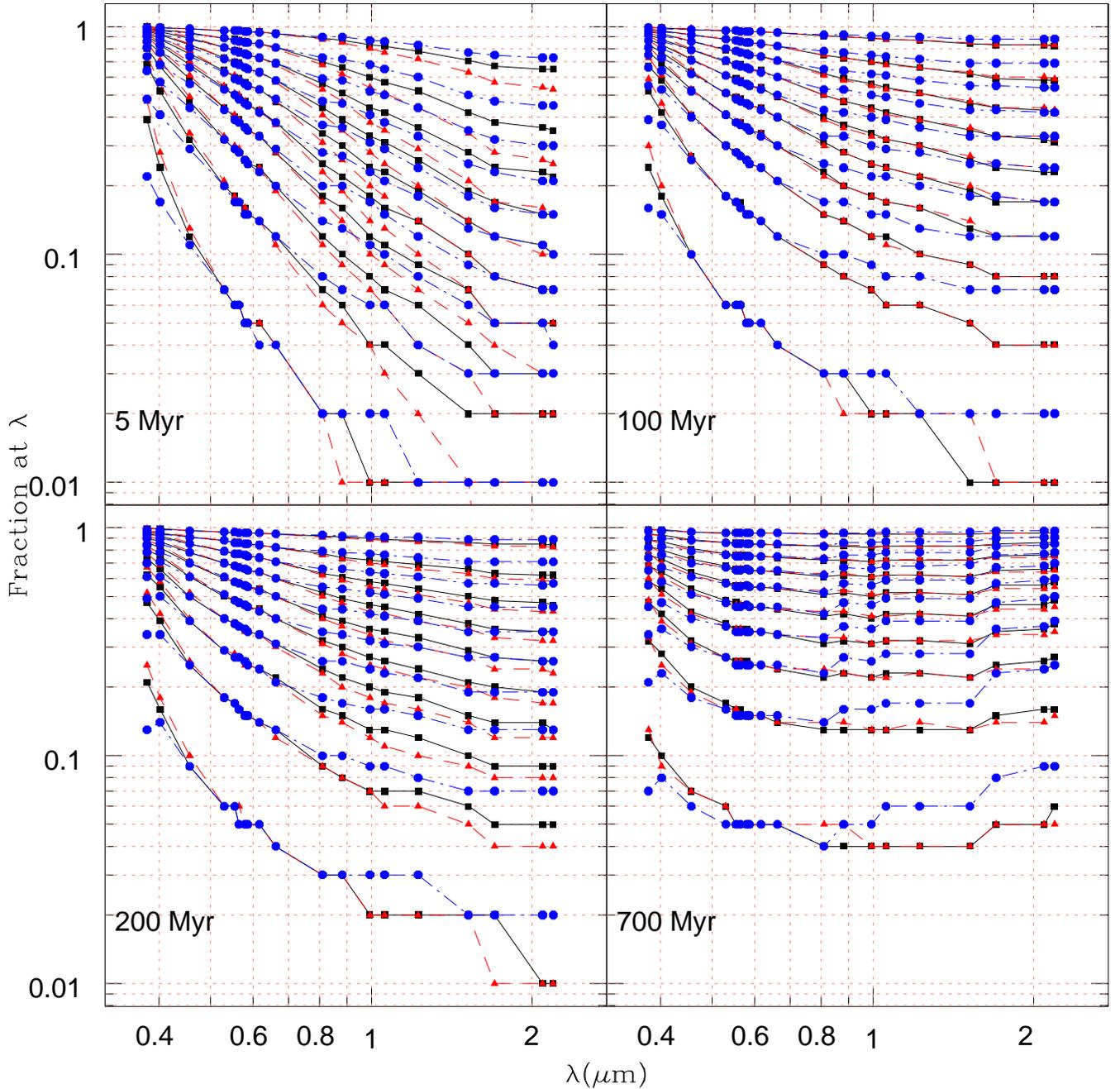}
\caption{Metallicity effect for the normalisation point at 5870\AA. Boxes represent solar metallicity, circles $\frac{1}{50} Z\odot$ and triangles are 2\,$Z\odot$.  We use the 
M05 models as reference.}
\label{metalopt}
\end{figure*}

\begin{figure*}
\centering
\includegraphics[scale=0.95]{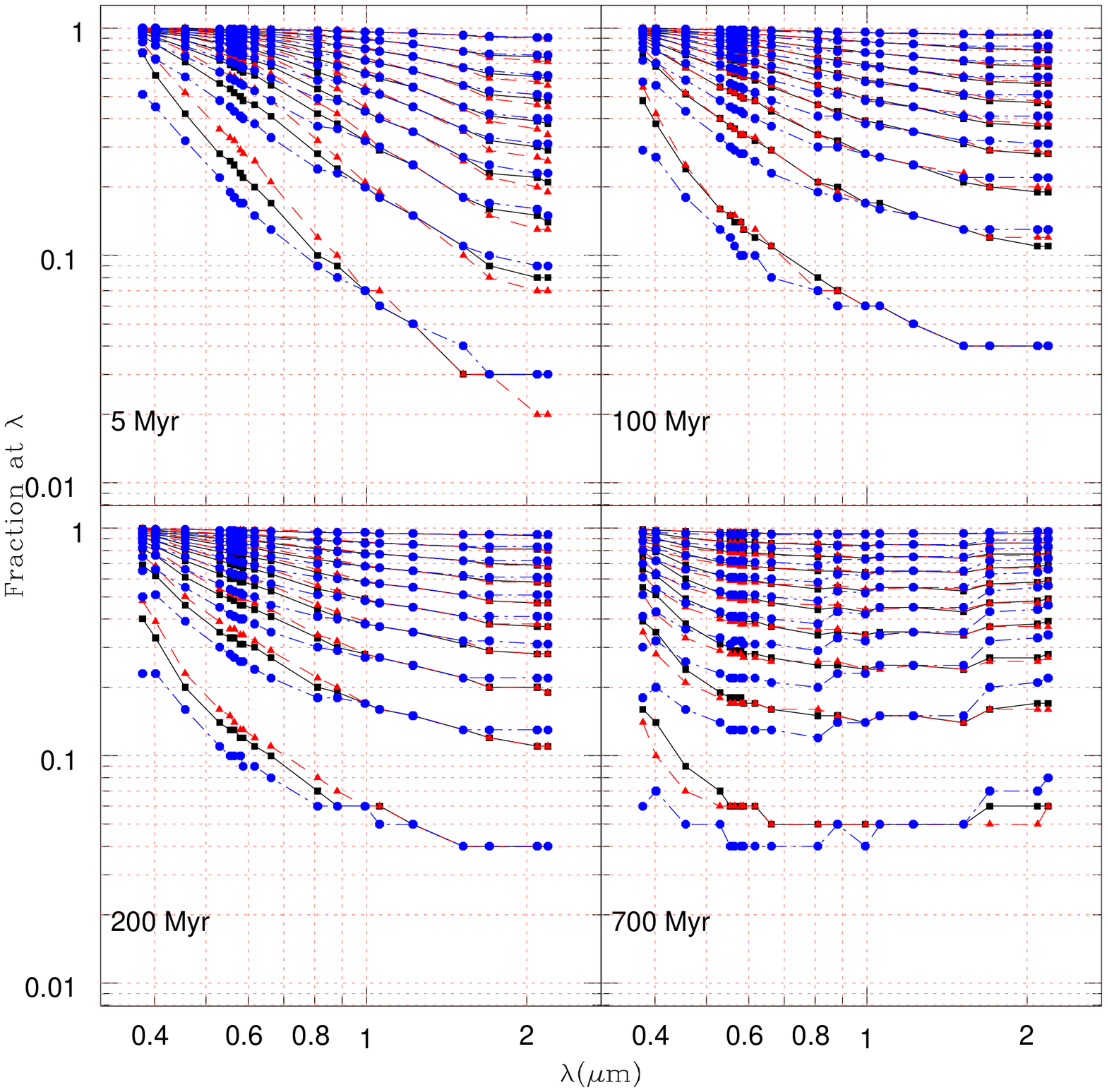}
\caption{Same as Fig.~\ref{metalopt} for the normalisation point at 1.223\mc.}
\label{metalnir}
\end{figure*}

Another important ingredient in stellar population fitting is the metallicities used. As 
shown by \citet{chen10}, the results of the fitting have a weaker dependence on metallicity than age. The question which arises here is, does 
metallicity affect the propagation of the averaged stellar populations? We investigate this effect with M05 SSPs with 3 different metallicities 
($\frac{1}{50}Z\odot$; $Z\odot$; and 2\,$Z\odot$) and the same age grid as in Fig.~\ref{fractionsOPT}. The results are shown 
in Figs.~\ref{metalopt} and \ref{metalnir}. Clearly the propagation of the contributions has a negligible dependence on metallicity. Thus, 
one can use the condensed population vectors proposed by \citet{cid04,cid05} to propagate the fitting results over all $\lambda$s.

\begin{figure*}
\centering
\includegraphics[scale=0.95]{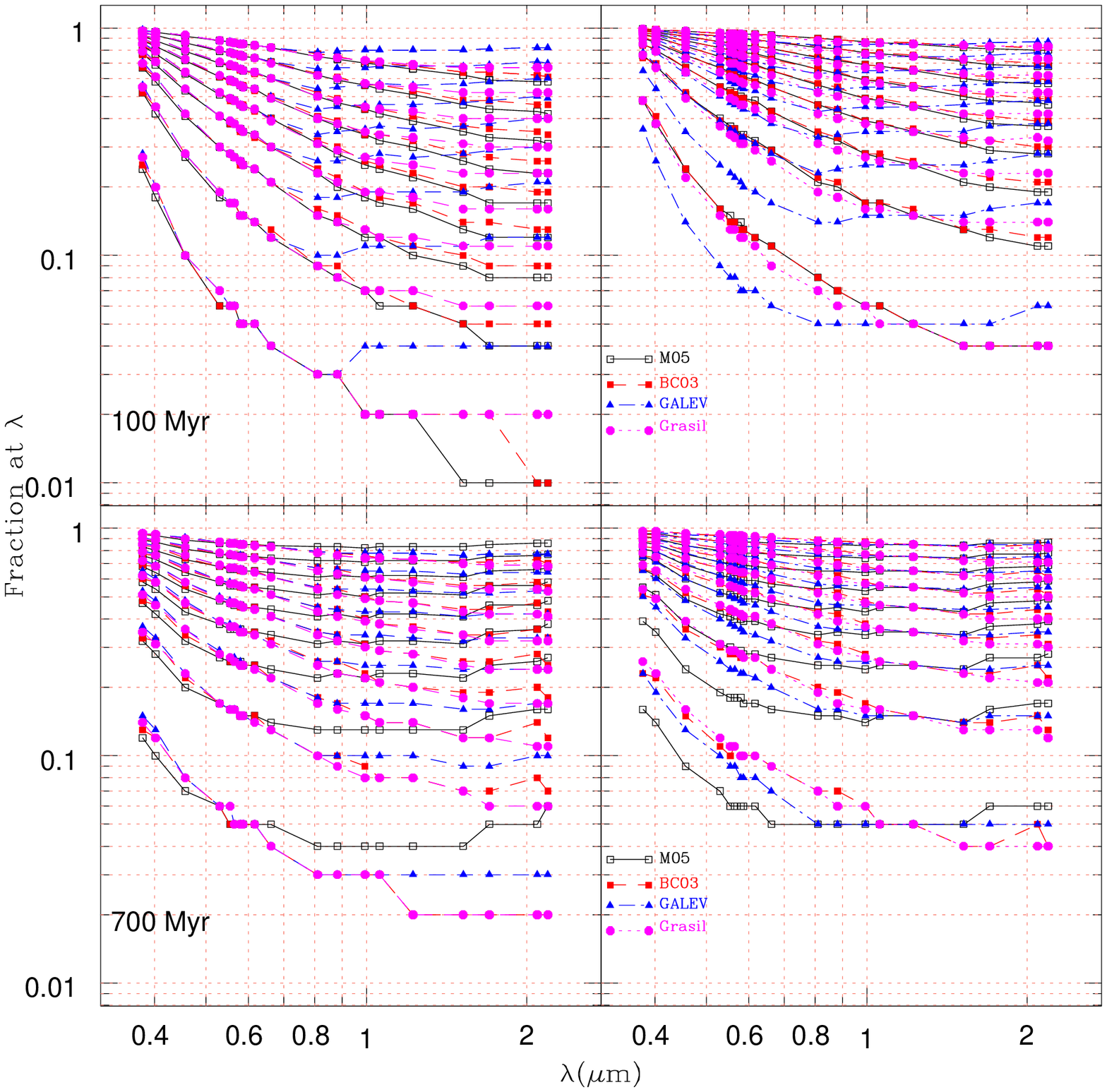}
\caption{Comparison of different EPS models. Normalised at 5870\AA\ (left) and 1.223\mc (right). Note that we 
use 100 Myr + 13\,Gyr SSPs.}
\label{comparamodels}
\end{figure*}

All the tests described above were made using the M05 models, but, as stated in Sec.~\ref{epsmodels}, there are more EPS models available in the 
literature covering simultaneously the spectral region between $\sim$3500\AA\ and 2.5\mc. Thus, it is necessary to test if the selection of EPS models 
will produce different results in the propagation of the synthesis 
results over different $\lambda$s. In Fig.~\ref{comparamodels} we compare the different models among each other. It is clear that in the case of 
the optical normalisation point (5870\AA), the four models produce very similar results in the interval between 3800\,\AA\ and 9000\,\AA, but a discrepancy 
between {\sc galev}/M05 and {\sc grasil}/BC03 models is observed in the NIR. Such a discrepancy is due to the well known fact that {\sc galev} and M05
models do include stars in the TP-AGB phase (see M05, for example), which is more sensitive to the NIR than the optical, i.e TP-AGB stars account for 25 to 40\% of 
the bolometric light of an SSP, and for 40 to 60\% of the light emitted in the K-band \citep[see][and references therein]{schulz02,maraston05}. 
However, there is a difference between {\sc galev} and M05 models, enhanced in the 100 Myr population. 
There are two possible explanations for this discrepancy: one is associated with the different onset age of the TP-AGB on the models.
A high TP-AGB contribution at 100\,Myr, as applied by \citet{schulz02}, which is excessively high when compared to young  Large Magellanic Cloud 
globular clusters \citep{maraston98,marigo08}. 
The other is associated with the way in which the TP-AGB treatment is made \citep{maraston06,bruzual07}. {\sc galev} includes TP-AGB by means of 
isochrones \citep[Padova94 + improved TP-AGB models][]{bertelli94,girardi00,marigo08}, 
while M05 is based on a different approach, the fuel consumption theorem.  According to \citet{maraston05}, the stellar 
luminosity during the evolutionary phases that follow or suffer from mass loss cannot be predicted by stellar tracks, because 
there is no theory linking mass-loss rates to the basic stellar parameters, such as luminosity. 

Note that the trend observed in Fig.~\ref{comparamodels} can also be associated with the fact that in M05 models, the TP-AGB 
contributes with 40\% to the bolometric flux, and 80\% to the K-band. This is higher than the ``simpler" calculations (made by the Padova group at the time) used by \citet{schulz02}, in which only some thermal pulses
have been included \citep{girardi00}.

\begin{figure*}
\begin{minipage}[b]{\linewidth}
\includegraphics[scale=0.95]{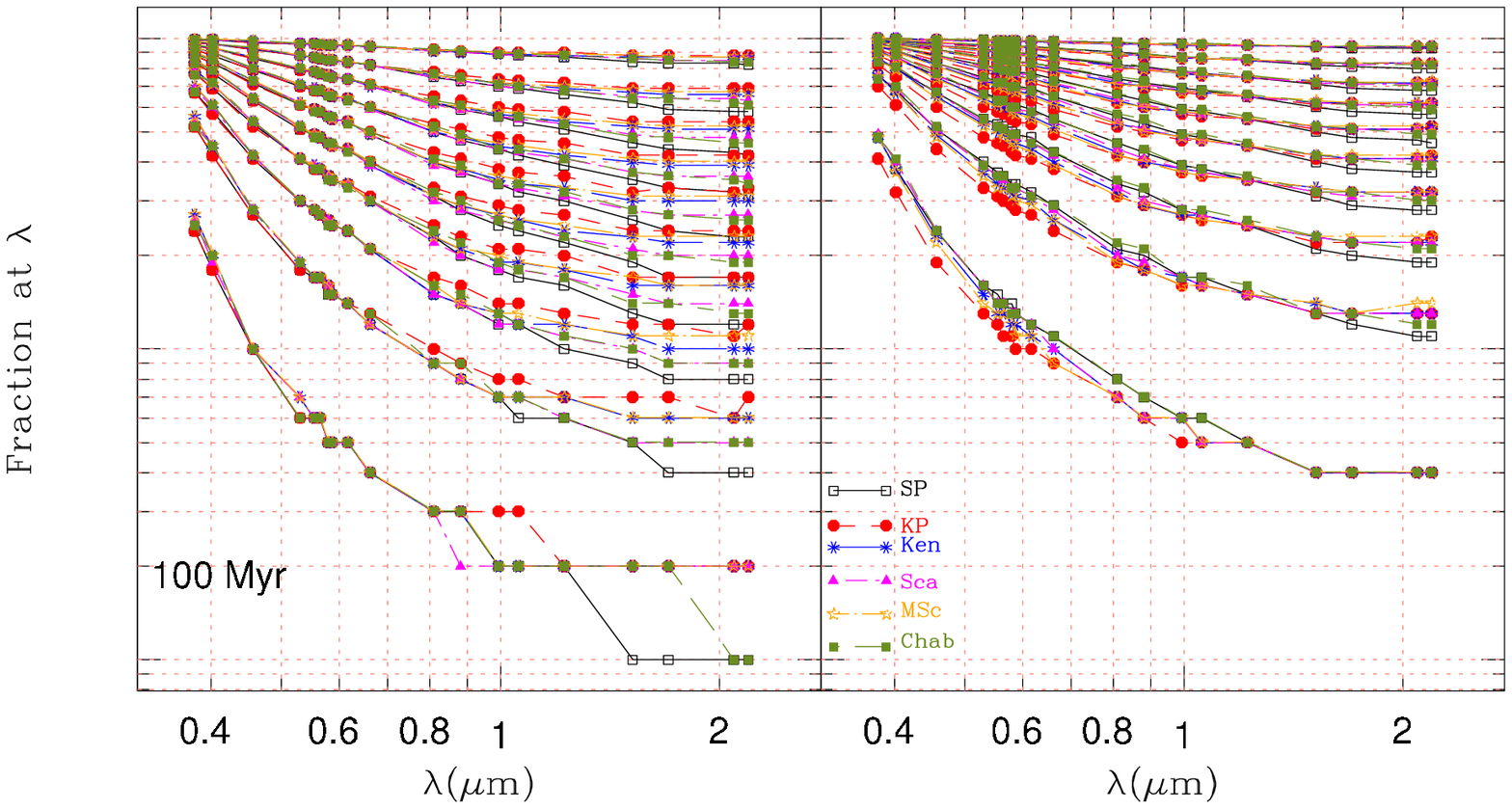}
\caption{Effect of different IMFs. At left the normalisation point is 5870\AA\ and at right 1.223\mc. Salpeter (SP) and Kroupa (KP) where taken from M05 models; Kennicutt (Ken),
Scalo (Sca) and Miller \& Scalo (MSc) IMFs were taken from {\sc grasil} EPS and Chabrier (Chab) are from {\sc galaxev}.}
\label{imf}
\end{minipage}
\hfill
\begin{minipage}[b]{\linewidth}
\includegraphics[scale=0.95]{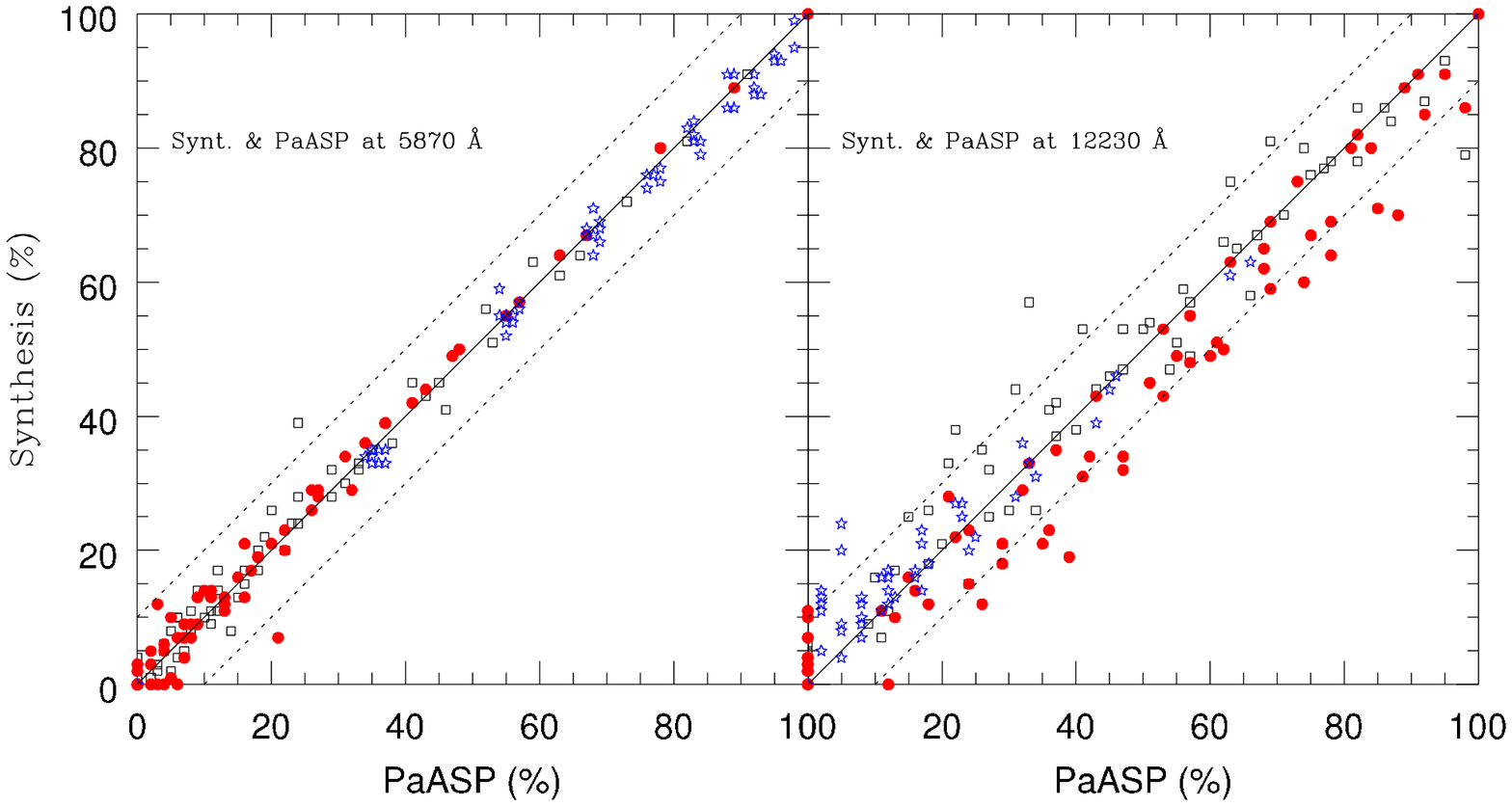}
\caption{\paasp\ tests. Stars, boxes and circles represent the 5\,Myr, 700\,Myr and 13 Gyr SSPs, respectively. 
Identity is shown by the solid line and dotted lines indicate the 90\% confidence level.}
\label{paasp}
\end{minipage}
\hfill

\end{figure*}

In addition, the tests were performed with the \citet{salpeter55} IMF. To see the effect of the IMF on the averaged stellar populations over
all $\lambda$s, we repeat the same exercises for different IMFs and summarise the results in Fig.~\ref{imf}. This figure suggests that 
the IMF does not play an important role in the synthesis propagation to other spectral regions.

\subsection{Testing \paasp}

In order to test if we are able to predict the averaged stellar populations over all lambdas, 
extensive simulations were performed to evaluate the \paasp's ability to recover input parameters. 
In short, we build artificial galaxy spectra by mixing \citet{maraston05} solar metallicity SSP models of 
 5\,Myr, 700\,Myr and 13\,Gyr in different proportions, and perform stellar population 
synthesis with normalisation points at the optical (5870\AA) and NIR (12230\AA). To this purpose we 
use the {\sc starlight} code optimised to the optical region \citep{cid04,cid05,mateus06,asari07,cid08}, and 
which also produces reliable results when applied to the NIR spectral \citep{riffel09,riffel10}. 
 
In summary, {\sc starlight} fits an observed spectum $O_{\lambda}$ with a combination, in different proportions, of
$N_{*}$ SSPs.   Basically, it solves the following equation \citep{cid05}:
\begin{equation}
M_{\lambda}=M_{\lambda 0}\left[\sum_{j=1}^{N_{*}}x_j\,b_{j,\lambda}\,r_{\lambda} \right] \otimes G(v_{*},\sigma_{*}),
\end{equation}
where  $M_{\lambda}$ is a model spectrum, $b_{j,\lambda}\,r_{\lambda}$ is the reddened spectrum of the $j$th SSP normalised at
$\lambda_0$; $r_{\lambda}=10^{-0.4(A_{\lambda}-A_{\lambda 0})}$ is the reddening term; $M_{\lambda 0}$ is the
synthetic flux at the normalisation wavelength; $\vec{x}$ is the population vector; $\otimes$ denotes the convolution
operator, and $G(v_{*},\sigma_{*})$ is the Gaussian distribution used to model the line-of-sight stellar
motions, which is centred at velocity $v_{*}$  with dispersion  $\sigma_{*}$.
The final fit is carried out with a simulated annealing plus Metropolis scheme, which searches for the
minimum of the equation:

\begin{equation}
\chi^2 = \sum_{\lambda}[(O_{\lambda}-M_{\lambda})w_{\lambda}]^2,
\end{equation}
where emission lines and spurious features are masked out by fixing $w_{\lambda}$=0. For a detailed description of {\sc starlight} see \citet{cid04,cid05}.

As base set we take \citet{maraston05} SSPs covering 14 ages, $t$= 0.001, 0.005, 0.01, 0.03, 0.05, 
0.1, 0.2, 0.5, 0.7, 1, 2, 5, 9, 13\,Gyr, and 4 metallicities, namely: 
$Z$= 0.02\,$Z_\odot$, 0.5\,$Z_\odot$, 1\,$Z_\odot$ and 2\,$Z_\odot$, summing up 56 elements.

To compare {\sc starlight} with \paasp\ predictions, we  
use the condensed population vector, which is obtained by binning
the synthesis results into  young, $x_Y$ ($t\leq \rm 5\times 10^7$yr);  intermediate-age, 
$x_I$ ($1\times 10^8 \leq t\leq \rm 2\times 10^9$yr) and old, $x_O$ ($t > \rm 2\times 10^9$yr) 
components \citep{riffel09,cid04}. These components were then taken to represent  
the 5\,Myr, 700\,Myr and 13\,Gyr old populations. These vectors are compared with the \paasp\ predictions 
in Fig.~\ref{paasp}.  

Clearly, our predictions are consistent with the stellar population 
synthesis, especially at the optical region, where the confidence level between predictions and  
synthesis is $\sim$95\% (i.e. almost all points fall in a region less than 5\% from the identity line). 
The confidence level drops to $\sim$85\% in the NIR.

\section{Final Remarks}\label{finalremarks}

We study the panchromatic stellar population components over the 3500\AA\ to 2.5\mc\ 
spectral region. {\bf In particular, we analyse how the spectral fitting of galaxies
based on light-fractions derived in a given spectral range can be propagated over 
all $\lambda$s. Dependencies on EPS models, age, metallicity, and stellar evolution 
tracks of four widely used EPS models ({\sc grasil}, {\sc galev}, Maraston and 
{\sc galaxev}) were taken into account.} Our main results are:

\begin{itemize}

\item The young ($t \lesssim$ 400 Myr) stellar population fractions derived in the optical cannot be directly 
compared to those derived in the NIR, and vice versa. For example, a contribution of $\sim$ 80\% of 
a 5\,Myr population at 3800\AA\ translates into only 5\% at 1.223\mc.

\item The intermediate to old age components ($t \gtrsim$ 500 Myr) can be directly compared from 
the NUV up to the NIR.

\item The metallicity dependence on the propagation of the stellar population fractions derived from 
NUV --- NIR is negligible.

\item Different EPS models produce similar results in the propagation of the synthesis 
results over the interval between 3800 \AA\ and 9000 \AA. However, a discrepancy 
between {\sc galev}/M05 and {\sc grasil}/BC03 models occurs in the NIR. Such a discrepancy may be due to the fact  that {\sc galev} and M05
models do include stars in the TP-AGB phase.

\item There is a difference between {\sc galev} and M05 models, enhanced in the 100 Myr population. Such a 
discrepancy may be associated with the way in which the TP-AGB treatment is made \citep{maraston06,bruzual07}. 

\item We test the effect of 6 different IMFs, and we conclude that the IMF is not important in the propagation of the synthesis 
results.

\item Extensive simulations were performed to evaluate \paasp's ability to recover input parameters. Our predictions are 
consistent with the stellar population synthesis with a confidence level between the predictions and  synthesis of $\sim$95\% in the 
optical and $\sim$85\% in the NIR. 

{\bf In summary, spectral fitting} results should not be directly propagated from the NIR to the NUV/Optical, or vice versa. Instead,
Eqs.~\ref{eqfrac} to ~\ref{eqfrac3} should be used for this purpose. However, since this is 
hard to do when dealing with large samples of objects,  we have created an on-line form, 
the {\bf Pa}nchromatic {\bf A}veraged {\bf S}tellar {\bf P}opulation - \paasp, available at: http://www.if.ufrgs.br/$\sim$riffel/software.html.
We also make available for download the tables with the results of the above equations for a wide range of ages (see Appendix ~\ref{appen}).

\end{itemize}

\section*{Acknowledgements}
{\bf We thank an  anonymous referee for interesting comments.}
R. R. thanks to the Brazilian funding agency CAPES. The {\sc starlight} project 
is supported by the Brazilian agencies CNPq, CAPES and
FAPESP and by the France-Brazil CAPES/Cofecub program.

\appendix

\section{How to use the \paasp\ form}\label{appen}

In this appendix we provide details how to use the \paasp\ form. \paasp\ is available at 
http://www.if.ufrgs.br/$\sim$riffel/software.html.

\begin{enumerate}

\item Bin your synthesis results ($\vec{x_j}$) into 3 population vectors: young, intermediate, and old ages. For more information on 
the definition of the populations vectors see \citet{cid04,cid05} and \citet{riffel09}.

\item Select the representative age of each vector, for example: 5\,Myr for young, 200\,Myr for intermediate, 
and 13\,Gyr for old. Tip: take as the representative population the ages of $\vec{x_j}$ with the largest contribution in each bin.

\item Put the percentage contribution (only integers and sum equal to 100\% are allowed) of each age
in the form, select the normalisation point and submit it.

\item You will be redirected to the results query page. There your inputs are marked in red
and the propagated results over all $\lambda$s (see text) are shown. You can also download the 
file with a table with propagation results for all the results of Eq. \ref{eqfrac3}, in fractions of 1\% for your query.

\item The 3 first columns of the table are the input fractions for a given normalisation 
point and for 3 ages, which are specified in the header (first and second lines). The next columns are the propagation of the 
results for all $\lambda$s.  Note that the results are given in 3 lines blocks: young, intermediate and old fractions, respectively.

\paasp\  is freely distributed and is supported by Brazilian funding agencies CAPES and CNPq. An acknowledgement for the use would be appreciated. 

\end{enumerate}

\end{document}